# Optical nano-imaging of gate-tuneable graphene plasmons

Jianing Chen[*,5,1], Michela Badioli[*,2], Pablo Alonso-González[*,1], Suko Thongrattanasiri[*,3], Florian Huth[*,1,6], Johann Osmond[2], Marko Spasenović[2], Alba Centeno[7], Amaia Pesquera[7], Philippe Godignon[8], Amaia Zurutuza[7], Nicolas Camara[9], Javier Garcia de Abajo[•,3], Rainer Hillenbrand[•,1,4], Frank Koppens[•,2]

1. CIC nanoGUNE, 20018 Donostia – San Sebastián, Spain
2. ICFO-Institut de Ciéncies Fotoniques, Mediterranean Technology Park, 08860 Castelldefels (Barcelona), Spain
3. IQFR-CSIC, Serrano 119, 28006 Madrid, Spain
4. IKERBASQUE, Basque Foundation for Science, 48011 Bilbao, Spain
5. Centro de Fisica de Materiales (CSIC-UPV/EHU) and Donostia International Physics Center (DIPC), 20018 Donostia-San Sebastián, Spain
6. Neaspec GmbH, 82152 Martinsried Munich, Germany
7. Graphenea S.A. 20018 Donostia - San Sebastián, Spain
8. CNM-IMB-CSIC–Campus UAB 08193 Bellaterra, Barcelona, Spain
9. GREMAN, UMR 7347, Université de Tours/CNRS, France
* These authors contributed equally to this work
• Corresponding authors: J.G.deAbajo@csic.es, r.hillenbrand@nanogune.eu, frank.koppens@icfo.es

**The ability to manipulate optical fields and the energy flow of light is central to modern information and communication technologies, as well as quantum information processing schemes. However, as photons do not possess charge, controlling them efficiently by electrical means has so far proved elusive. A promising way to achieve electric control of light could be through plasmon polaritons - coupled excitations of photons and charge carriers – in graphene[1-5]. In this two-dimensional sheet of carbon atoms[6], it is expected that plasmon polaritons and their associated optical fields can be readily tuned electrically by varying the graphene carrier density. While optical graphene plasmon resonances have recently been investigated spectroscopically[7,8], no experiments so far have directly resolved propagating plasmons in real space. Here, we launch and detect propagating optical plasmons in tapered graphene nanostructures using near-field scattering microscopy with infrared excitation light[9-11]. We provide real-space images of plasmonic field profiles and find that the extracted plasmon wavelength is remarkably short - over 40 times smaller than the wavelength of illumination. We exploit this strong optical field confinement to turn a graphene nanostructure into a tunable resonant plasmonic cavity with extremely small mode volume. The cavity resonance is controlled in-situ by gating the graphene, and in particular, complete switching on and off of the plasmon modes is demonstrated, thus paving the way towards graphene-based optical transistors. This successful alliance between nanoelectronics and nano-optics enables the development of unprecedented active subwavelength-scale optics and a plethora of novel nano-optoelectronic devices and functionalities, such as tunable metamaterials[12], nanoscale optical processing and strongly enhanced light-matter interactions for quantum devices[13] and (bio)sensing applications.**

Surface plasmons are electromagnetic waves that propagate along the surface of a metal[14]. Similar propagating waves are expected for graphene[4]. In fact, due to the 2D nature of the collective excitations in this material, the confinement of graphene plasmons is expected to be much stronger than that of metallic surface plasmons. However, launching and detecting graphene plasmons has so far remained a challenge: because of the large wave vector mismatch of graphene plasmons compared to free-space photons, plasmon excitation and detection by light is very inefficient. The first reports on graphene plasmon resonances were based on electron spectroscopies (inelastic electron scattering[15-17] and photoemission spectroscopy[18]), used to spectrally probe broad plasmons in large-area graphene. Recently, resonant coupling of propagating THz waves to plasmons in micro-ribbons[7] has been demonstrated, while IR near-field microscopy has been applied to observe the coupling of graphene plasmons to phonons[8]. These pioneering works have revealed the interaction between low-energy photons and graphene plasmons in the spectral domain. However, high-resolution nanoscale real-space imaging of the plasmonic modes is of fundamental importance to conclusively unveil propagating and localized plasmons in graphene sheets and nanostructures.

Here, we visualize for the first time graphene plasmons in real space. By employing scattering-type near-field microscopy (s-SNOM, from Neaspec GmbH) we excite and spatially image propagating and localized plasmons in tapered graphene ribbons at infrared frequencies. To that end, we scan the metalized tip of the s-SNOM over the sample while illuminating both tip and sample with infrared laser light. The tip acts as an optical antenna that converts the incident light into a localized near field below the tip apex[19]. The nanoscale field concentration provides the required momentum[11,20] for launching plasmons on graphene, as illustrated in Fig. 1a. Plasmon reflection at the graphene edges produces plasmon interference, which is recorded by collecting the light elastically scattered by the tip via far-field pseudo-heterodyne interferometry[21]. The detected signal as function of tip position yields a spatially resolved near-field image with nanometer-scale resolution.

A representative near-field image is shown in Fig. 1b, where the tip is scanned over a tapered graphene ribbon on 6H-SiC C-face[22], illuminated by infrared excitation light with a free-space wavelength $\lambda_0$ = 9.7 µm. One of the most distinct feature in this image is the presence of fringes parallel to the edge of the ribbon in its wider part. The distance between fringe maxima is approximately constant at ~130 nm inside the ribbon. We interpret these fringes as follows: the tip launches radial surface waves that propagate along the surface and reflect at the edges, partially reaching the tip again. Consequently, we probe the interference of forward- and backward-propagating plasmons. Within this basic physical picture, the maxima are separated by half the plasmon wavelength $\lambda_p/2$. Thus, we experimentally find a plasmon wavelength $\lambda_p$=260 nm, which is about a factor of 40 smaller than the free-space excitation wavelength. As we discuss further below, the near-field images represent the local density of optical states (LDOS). The calculated LDOS for a tapered ribbon is shown in Fig. 1c, matching very closely the experimental results (Fig. 1b), including the features at the more narrow part of the ribbon.

Our observation of a remarkably strong reduction in the electromagnetic field propagation wavelength $\lambda_p = \lambda_0/40$ can directly be attributed to the two-dimensionality and the unique conductance properties of graphene. Namely, the plasmonic properties of graphene are related to the optical conductivity of graphene[1,23], σ (e.g., $2\pi/\lambda_p$ ~$(\epsilon_r+1)\omega/4\pi \text{Im}\{\sigma\}$, where $\epsilon_r$ is the substrate permittivity and ω is the frequency). For sufficiently high doping, quantified through a Fermi energy $E_F$ exceeding the plasmon energy $E_p$, this yields[4,13]

$$\lambda_p = \lambda_0 \alpha \frac{E_F}{E_p} \frac{4}{\epsilon_r + 1} \quad \text{(Equation 1)}$$

Interestingly, this simplified equation reveals a relation between the plasmon wavelength and the free-space wavelength governed by the fine-structure constant α≈ 1/137. The observed $\lambda_p$=260 nm is in good agreement with the theoretical prediction of equation (1) for the specific substrate ($\epsilon_{r,SiC}$ = 1.6 for $\lambda_0$=9.7μm), which yields $\lambda_p$=260 nm assuming $E_F$~0.32eV. This value is about a factor of two higher than the intrinsic substrate-induced doping found in earlier studies of graphene on 6H-SiC C-face[24]. We speculate that narrow ribbons exhibit larger carrier densities, and this will be addressed in future studies.

In Fig. 2, we present a more detailed experimental study of the plasmon properties in graphene nano-structures by taking advantage of the strong dependence of the dielectric constant of the SiC substrate $\epsilon_{r,SiC}$ on the excitation wavelength[25]. This allows us to tune the plasmon wavelength over a wide spectral range by just slightly changing the excitation wavelength, as the plasmon wavelength strongly depends on the dielectric constant of the substrate. The near-field images of relatively wide ribbons are displayed in Fig. 2a, showing that the spacing of the interference fringes significantly decreases with increasing $\epsilon_{r,SiC}$. This observation is qualitatively consistent with equation (1), as a larger substrate permittivity yields a smaller graphene plasmon wavelength. Quantitatively, we obtain good agreement between the plasmon wavelengths extracted from the near-field images (Fig. 3a, symbols) and the prediction of equation (1) for graphene on SiC (Fig. 3a, solid curves), using literature values for the dielectric constant of SiC[25] and an intrinsic doping $E_F$=0.4 eV.

Our experimental observation of an extremely short plasmon wavelength compared to the excitation wavelength comes along with an extraordinary confinement of the infrared field perpendicular to the graphene sheet, characterized by a decay length $\delta \approx \lambda_p/2\pi$[11]. This means that narrow graphene ribbons are ideally suited to confine light down to extremely small volumes. In Fig. 2b, we show near-field images of the tapered ribbons where the width $W$ reaches values smaller than the plasmon wavelength $\lambda_p$. These images clearly reveal two distinct localized modes (indicated by red and white arrows) which coexist with a resonant enhancement of the near-field signal, comparable to the observations in Ref.[32]. The resonance condition depends on $\lambda_p$ and the ribbon width $W$, as we observe a clear shift of the localized modes to a wider part of the ribbons for increasing $\lambda_p$. The width $W$ for which these two modes occur, normalized to the plasmon wavelength $\lambda_p$, is shown in Fig. 3b, from which we extract the resonance conditions $W$~$0.3\lambda_p$ and ~$0.6\lambda_p$.

To obtain a better understanding of the physical mechanisms that underlie the observation of these resonant optical modes, as well as the interference fringes, we use a numerical model to calculate the field backscattered by the tip. To that end, we describe the microscope tip (on average 60 nm away from the surface) as a vertically oriented point dipole[11] that couples efficiently to propagating and localized plasmon modes[26,27]. The tip launches plasmons that are reflected at the ribbon edges. These plasmons act back on the tip, and are subsequently scattered into photons, which we detect. The detected signal can be interpreted as a probe for the vertical component of the LDOS. In order to simulate two-dimensional LDOS maps for a dipole 60 nm above a tapered graphene ribbon, we combine one-dimensional LDOS profiles of graphene ribbons of fixed width. We justify this approach because the ribbon width along the graphene triangle varies adiabatically and plasmon reflection at the tip of the triangle is expected to be small.

The LDOS maps calculated for two different values of the substrate permittivity $\varepsilon_{r,SiC}$ are shown in Figs. 1c and 3c. As in the experimental s-SNOM images, the LDOS maps reveal interference fringes and localized modes near the tip of the ribbon. The fringe spacing matches quantitatively the experimental results and the spacing increases with decreasing $\varepsilon_r$, associated with an increase in $\lambda_p$, as predicted by equation 1. The good agreement between experiment and theory confirms that the fringes in the wider part of the ribbon are due to plasmon interference caused by plasmon reflections at the graphene edges. We remark that both the LDOS and the experimental images exhibit their maximum away from the graphene edge, and that the fringe spacing slightly increases closer to the edge. This can be explained by the electromagnetic boundary conditions at the edges (further discussed below) and the fact that the plasmon wave vector perpendicular to the edge does not have a single value but rather a finite distribution around $2\pi/\lambda_p$ (SOM).

The comparison between the calculated LDOS maps and the experimental data in Fig. 2 can be used to estimate plasmon propagation distances. We observe five well-defined interference fringes away from a single edge. The fringes inside the ribbon decay due to the circular character of the plasmons and due to intrinsic losses. In addition, the peak close to the edge is relatively strong which we attribute to the strong concentration of electromagnetic field close to the edge. Both observations are consistent with our LDOS calculations for plasmon losses corresponding to a mobility of 1200 cm$^2$/Vs[4] (see SOM). This mobility is typical of similar graphene ribbons under ambient experimental conditions. In particular, the observation of the co-existence of strong reduction in plasmon wavelength (and thus strong optical field confinement) and relatively long propagation distance is very promising and a unique feature of plasmons carried by graphene. We emphasize that much longer propagation distances are expected for higher mobility graphene.

In our LDOS model interpretation, the localized modes near the tip of the graphene ribbon (marked by arrows) are explained as localized graphene plasmon resonances, which occur for specific values of the ribbon width ($W=0.37\lambda_p$ and $0.82\lambda_p$), where the strong concentration of the electromagnetic field yields an enhanced plasmon-dipole interaction,[13] and therefore, an increase in the near-field signal. Interestingly, for both theory and experiment, the profiles of the two localized modes are distinctly different from those of conventional Fabry-Perot cavity modes. For example, the lowest order mode (indicated by white arrows) exhibits field maxima at the graphene edges, while for a conventional lowest order Fabry-Perot mode the field is maximum in the middle. This is because graphene plasmons are being reflected at the boundaries with a reflection coefficient of approximately one (zero phase), rather than the coefficient of minus one (π phase) characteristic of the conventional Fabry-Perot model (see SOM).

One of the most appealing advantages of graphene plasmonics is the capability to control and switch nanoscale optical fields in-situ. Here, we demonstrate very effective electrical control of nanoscale optical fields by applying an electric field perpendicular to the graphene sheet, which allows for varying the carrier density in the ribbon. To this end, we have fabricated tapered ribbons based on CVD-grown graphene (source: graphenea S.A.) on a $SiO_2$ substrate with a Si backgate. By applying a backgate voltage $V_B$, we tune the carrier density and thus the Fermi energy $E_F \sim (V_B - V_D)^{1/2}$, where $V_D$ is the voltage that needs to be applied to offset the intrinsic doping, i.e., to reach the Dirac point; $V_D$ is extracted from optical measurements, as we discuss below). The effect of changing $V_B$ on the near-field images is shown in Fig. 4a, where the Fermi energy is tuned over a wide range from ≈0 to 0.15 eV. For $V_B-V_D>10$ V, the general near-field features are comparable to those of ribbons on SiC substrates, including the two local ribbon resonances indicated by white and red arrows. By increasing $V_B$, we find that the resonances (signal maxima) shift towards larger ribbon width, which we attribute to an increase in plasmon wavelength when the carrier density and thus also the Fermi energy increases (see Equation 1). The extracted value of $\lambda_p$ as a function of gate voltage is shown in Fig. 4b (red circles correspond to the tapered ribbon shown in Fig. 4a, and green crosses correspond to additional ribbons shown in the SOM). The calculated plasmon dispersion, represented by the blue contour plot in Fig. 4b, includes plasmon damping: for small carrier densities ($E_F \lesssim E_p$), inter-band carrier excitations can strongly damp the plasmons, but to first order these transitions are suppressed for $E_F \gtrsim E_p$. This is illustrated by the schematics in Fig. 4b. The data agree well with the calculated plasmon dispersion, by assuming either a relatively low $\varepsilon_r=1$ for the $SiO_2$ substrate (while $\varepsilon_r \sim 2$ is expected), or a factor two larger Fermi energy associated to a larger carrier concentration. We speculate that the lower predicted value for $\lambda_p$ is due to charge accumulation at the ribbon tip, as predicted in Ref[28]. Further understanding of the inhomogeneous charge distribution will require more detailed studies, which go beyond the scope of this work.

The effect of plasmon damping offers the intriguing capability to actively switch graphene plasmons on and off by electric fields. Experimentally, we clearly observe very strong plasmon damping in the left panel of Fig. 4a (corresponding to $E_F \lesssim E_p$), where the ribbon does not show any signal compared to the substrate. We illustrate electrostatic switching of graphene plasmons in more detail in Fig. 4c, which portraits line scans across a ribbon of width $W$=200 nm (vertical axis), while changing $V_B$ (horizontal axis). At the Dirac point ($V_B=V_D$), the near-field signal is dramatically depleted on the whole ribbon. With increasing Fermi energy, at both sides of the Dirac point, two fringes show up near the graphene edges, and for even higher positive $V_B$, the familiar localized mode emerges with a maximum in the center of the ribbon. The signals at both sides of the Dirac point are attributed to plasmons carried by either p- or n-type charge carriers, while for $E_F<E_{pl}$ the complete signal depletion is due to inter-band transitions. These conclusions are supported by calculated LDOS profiles as a function of Fermi energy (lower plot of Fig. 4c), in excellent agreement with the experimental observations.

In summary, here and in Ref.[32] electrical control of confined and propagating plasmons is demonstrated, thus providing an ingenious solution to a major problem in plasmonics, as it facilitates the design and miniaturization of active nanoscale photonic devices[29,30]. This leads to a new paradigm in optical and opto-electronic telecommunications and information processing. As an alternative to plasmon excitation and detection by (effective) dipoles, plasmons can also be resonantly excited by light in graphene nanocavities,[13], enabling strong enhancement of light absorption in graphene[31], and a new basis for infrared detectors and light-harvesting devices.

**Captions**

Fig 1: Imaging propagating and localized graphene plasmons by s-SNOM. a) Schematic of the experimental configuration used to launch and detect propagating surface waves in graphene. b) Near-field amplitude image acquired for a tapered graphene ribbon on top of 6H-SiC. The imaging wavelength is $\lambda_0$=9.7µm. The tapered ribbon is 12 µm long and up to 1 µm wide. c) Calculated local density of optical states (LDOS) at a distance of 60 nm from the graphene surface. Simulation parameters: graphene mobility µ=1000 cm$^2$/Vs and Fermi energy $E_F$=0.4 eV, substrate $\varepsilon_r$ =1.

Fig. 2: Controlling the plasmon wavelength over a wide range. (a) Near-field optical images of a tapered ribbon and (b) a ribbon of ~1 µm width (upper panels), both on the same 6H-SiC substrate. The topography (obtained by AFM) is shown in grayscale in the leftmost and rightmost panels, and outlined by dashed lines in the central panels. The line traces in the left and right panels are extracted from the near-field images for $\lambda_0$=9,200 nm and $\lambda_0$=10,152 nm. Red and white arrows indicate the resonant localized modes.

Fig 3: Comparison of theoretical model with experimental results. a) Experimentally extracted plasmon wavelength from interference fringes (blue crosses) and localized modes (red cross), compared to the calculated plasmon dispersion (blue curves, see SOM) for graphene with intrinsic doping of 0.2 and 0.4 eV on a SiC-6H substrate. For illustration, the dotted lines represent the plasmon dispersion for graphene on a substrate with fixed $\varepsilon_r$. b) Experimentally obtained resonance conditions $W/\lambda_p$ extracted from localized-mode measurements. Red crosses and black circles correspond to the modes indicated by red and white arrows in Fig. 2, respectively. (c,d) Spatial distribution of the LDOS calculated for homogeneous ribbons of increasing width (from bottom to top), supported on a dielectric with $\varepsilon_r$ =3 (left) or $\varepsilon_r$ =0.5 (right). The ribbon width of the two lowest-order modes is shown in units of the plasmon wavelength of extended graphene $\lambda_p$.

Fig. 4: Plasmonic switching and active control of the plasmon wavelength by electrical gating. a) Near-field amplitude images for tapered (CVD-grown) graphene ribbons on a Si/SiO$_2$ (300 nm) substrate, acquired while applying backgate voltages ranging from -15 to +11 V. We extract the Dirac voltage $V_D$ from optical images and by fitting the data to the model. Localized modes are indicated by white and red arrows. The illumination wavelength is $\lambda_0$=11.06 µm. b) Plasmon wavelength experimentally extracted from localized mode resonances indicated by red arrows. Red circles represent the datasets presented in a), while green markers correspond to one additional dataset presented in the SOM. The schematics illustrate plasmon damping by inter-band transitions. The calculated plasmon dispersion is represented through the colour map corresponding to the reflection coefficient including inter- and intra-band scattering processes (through the random-phase approximation). The dashed curve represents the calculated plasmon wavelength. c) Near-field amplitude signal for a ribbon with mode B appearing at $V_B$-$V_D$=35 V, using $\lambda_0$=10.6 µm. A full near-field image of this ribbon is shown in the SOM.

Upper panel: experiment. Bottom panel: calculated LDOS for fixed ribbon width and substrate $\varepsilon_r=1$.

## References

1. Wunsch, B., Stauber, T., Sols, F. & Guinea, F. Dynamical polarization of graphene at finite doping. *New Journal of Physics* **8**, 318–318 (2006).
2. Hwang, E. H. & Sarma, das, S. Dielectric function, screening, and plasmons in two-dimensional graphene. *Phys. Rev. B* **75**, (2007).
3. Polini, M. *et al.* Plasmons and the spectral function of graphene. *Phys. Rev. B* **77**, (2008).
4. Jablan, M., Buljan, H. & Soljačić, M. Plasmonics in graphene at infrared frequencies. *Phys. Rev. B* **80**, 245435 (2009).
5. Hill, A., Mikhailov, S. A. & Ziegler, K. Dielectric function and plasmons in graphene. *EPL (Europhysics Letters)* **87**, 27005 (2009).
6. Novoselov, K. S. *et al.* Electric field effect in atomically thin carbon films. *Science* **306**, 666–669 (2004).
7. Ju, L. *et al.* Graphene plasmonics for tunable terahertz metamaterials. *Nature Nanotech* **6**, 630–634 (2011).
8. Fei, Z. *et al.* Infrared Nanoscopy of Dirac Plasmons at the Graphene–SiO2 Interface. *Nano Lett* **11**, 4701–4705 (2011).
9. Keilmann, F. & Hillenbrand, R. Near-field microscopy by elastic light scattering from a tip. *Philos Transact A Math Phys Eng Sci* **362**, 787–805 (2004).
10. Huber, A., Ocelic, N., Kazantsev, D. & Hillenbrand, R. Near-field imaging of mid-infrared surface phonon polariton propagation. *Applied Physics Letters* **87**, 081103 (2005).
11. Novotny, L. & Hecht, B. *Principles of nano-optics*. (Cambridge University Press: 2006).
12. Vakil, A. & Engheta, N. Transformation Optics Using Graphene. *Science* **332**, 1291–1294 (2011).
13. Koppens, F. H. L., Chang, D. E. & García de Abajo, F. J. Graphene Plasmonics: A Platform for Strong Light–Matter Interactions. *Nano Lett* **11**, 3370–3377 (2011).
14. Raether, H. *Surface plasmons*. (Springer Berlin: 1988).
15. Liu, Y., Willis, R. & Emtsev, K. Plasmon dispersion and damping in electrically isolated two-dimensional charge sheets. *Phys. Rev. B* (2008).
16. Eberlein, T., Bangert, U., Nair, R., Jones, R. & Gass, M. Plasmon spectroscopy of free-standing graphene films. *Phys. Rev. B* (2008).
17. Zhou, W. *et al.* Atomically localized plasmon enhancement in monolayer graphene. *Nature Nanotech* (2012).doi:10.1038/nnano.2011.252
18. Bostwick, A., Ohta, T., Seyller, T., Horn, K. & Rotenberg, E. Quasiparticle dynamics in graphene. *Nat Phys* **3**, 36–40 (2006).
19. Hillenbrand, R., Taubner, T. & Keilmann, F. Phonon-enhanced light matter interaction at the nanometre scale. *Nature* **418**, 159–162 (2002).
20. Hecht, B., Bielefeldt, H., Novotny, L. & Inouye, Y. Local Excitation, Scattering, and Interference of Surface Plasmons. *Phys. Rev. Lett* (1996).
21. Ocelic, N., Huber, A. & Hillenbrand, R. Pseudoheterodyne detection for background-free near-field spectroscopy. *Applied Physics Letters* **89**, 101124 (2006).

22. Camara, N. *et al.* Current status of self-organized epitaxial graphene ribbons on the C face of 6H–SiC substrates. *J. Phys. D: Appl. Phys.* **43**, 374011 (2010).
23. Castro Neto, A. H., Peres, N. M. R., Novoselov, K. S. & Geim, A. K. The electronic properties of graphene. *Reviews of Modern Physics* **81**, 109–162 (2009).
24. Crassee, I. *et al.* Multicomponent magneto-optical conductivity of multilayer graphene on SiC. *Phys. Rev. B* **84**, (2011).
25. Hofmann, M., Zywietz, A. & Karch, K. Lattice dynamics of SiC polytypes within the bond-charge model. *Phys. Rev. B* (1994).
26. Nikitin, A. Y., Guinea, F., Garcia-Vidal, F. J. & Martín-Moreno, L. Edge and waveguide terahertz surface plasmon modes in graphene microribbons. *Phys. Rev. B* **84**, (2011).
27. Christensen, J., Manjavacas, A., Thongrattanasiri, S., Koppens, F. H. L. & García de Abajo, F. J. Graphene plasmon waveguiding and hybridization in individual and paired nanoribbons. *ACS Nano* **6**, 431–440 (2012).
28. Silvestrov, P. & Efetov, K. Charge accumulation at the boundaries of a graphene strip induced by a gate voltage: Electrostatic approach. *Phys. Rev. B* **77**, (2008).
29. Atwater, H. A. The Promise of Plasmonics. *Sci Am* **296**, 56–62 (2007).
30. Zia, R., Schuller, J. A., Chandran, A. & Brongersma, M. L. Plasmonics: the next chip-scale technology. *Materials Today* **9**, 20–27 (2006).
31. Thongrattanasiri, S., Koppens, F. & García de Abajo, F. Complete Optical Absorption in Periodically Patterned Graphene. *Phys. Rev. Lett* **108**, (2012).

32. Z. Fei, A. S. Rodin, G. O. Andreev, W. Bao, A. S. McLeod, M. Wagner, L. M. Zhang, Zeng Zhao, G. Dominguez, M. Thiemens, M. M. Fogler, A. H. Castro-Neto, C. N. Lau, F. Keilmann, D. N. Basov, "Gate-tuning of graphene plasmons revealed by infrared nano-imaging", (in preparation).

# Fig. 1

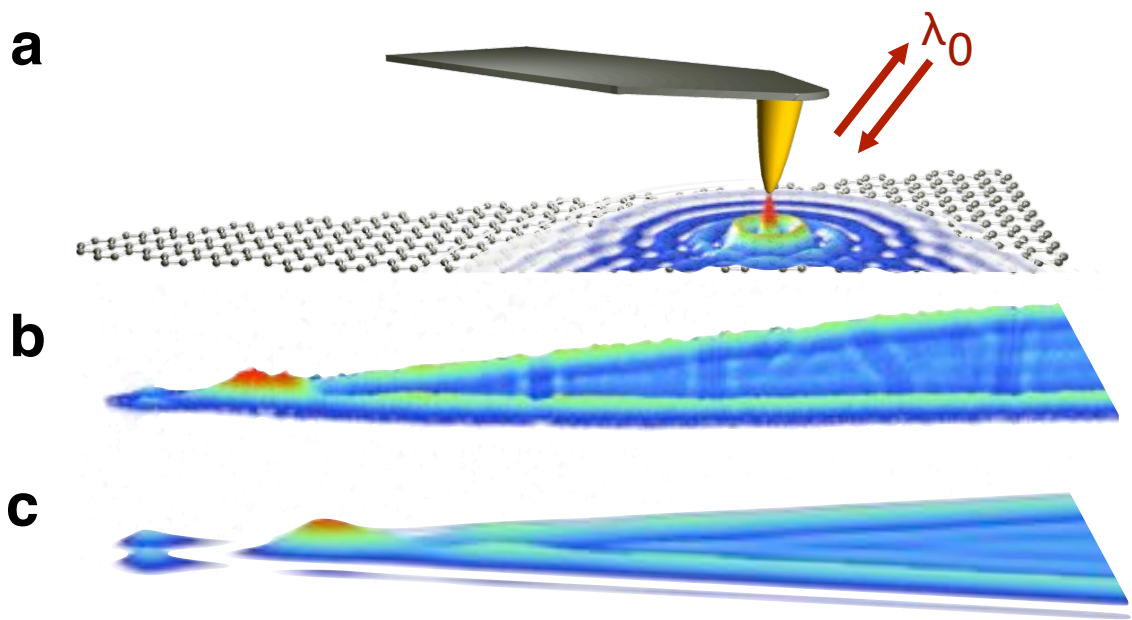

# Fig. 2

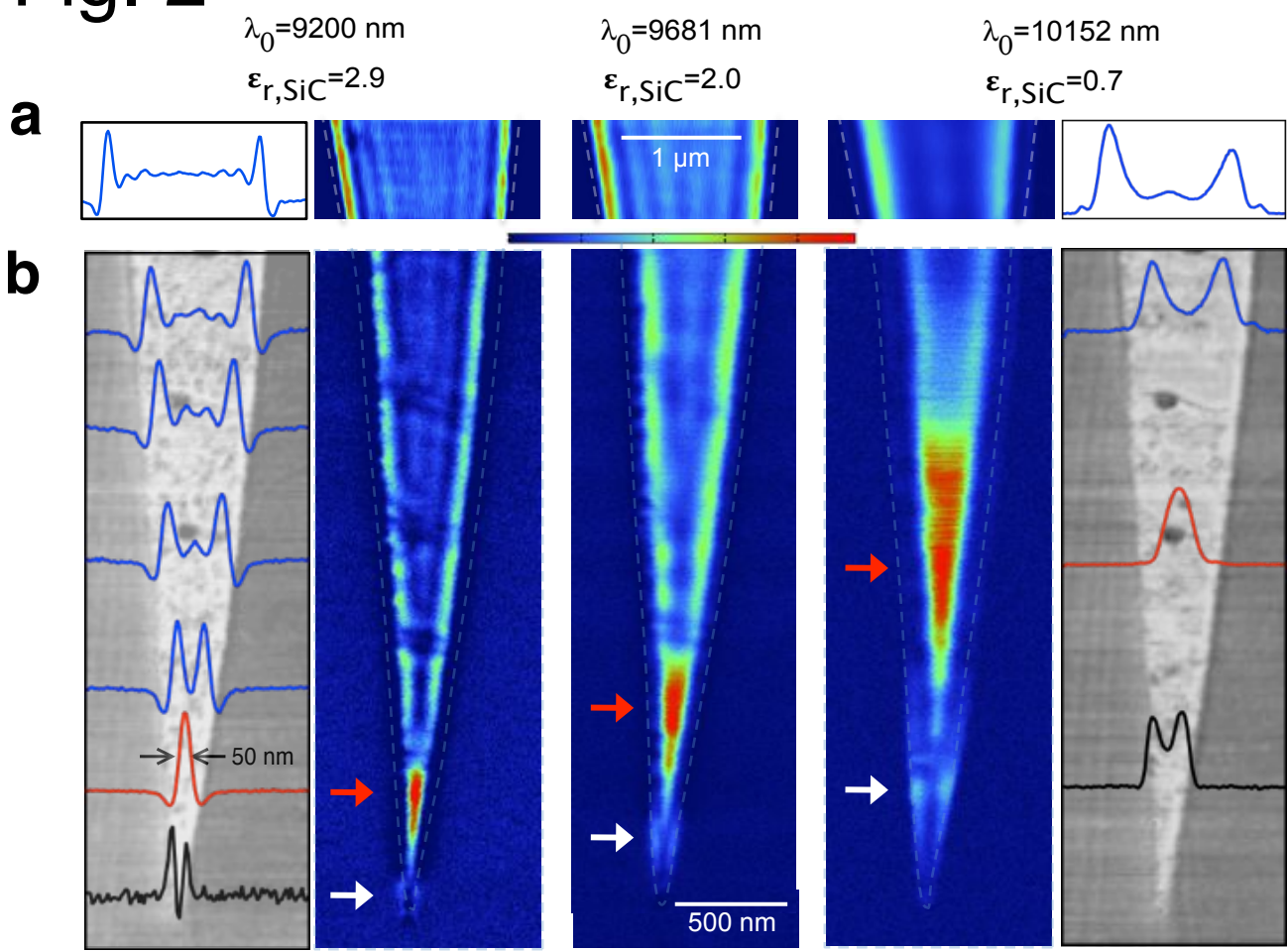

Fig. 3

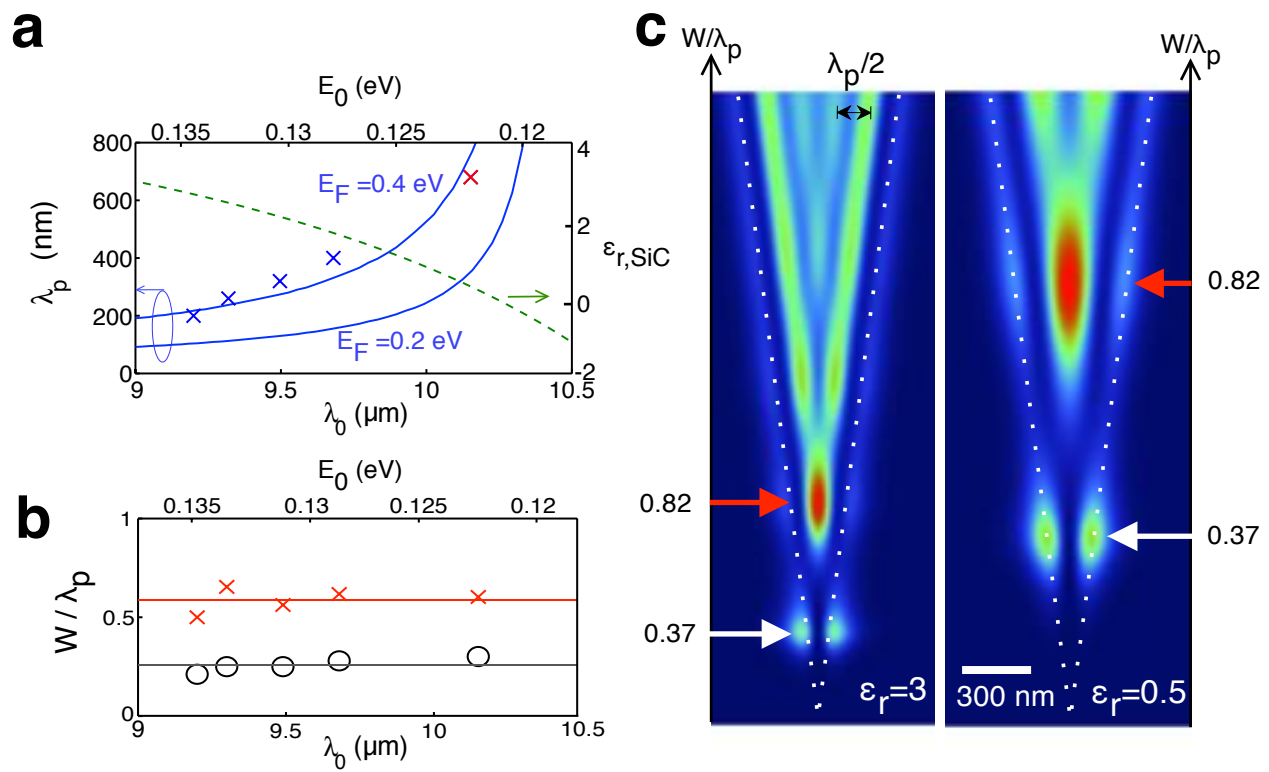

Fig. 4

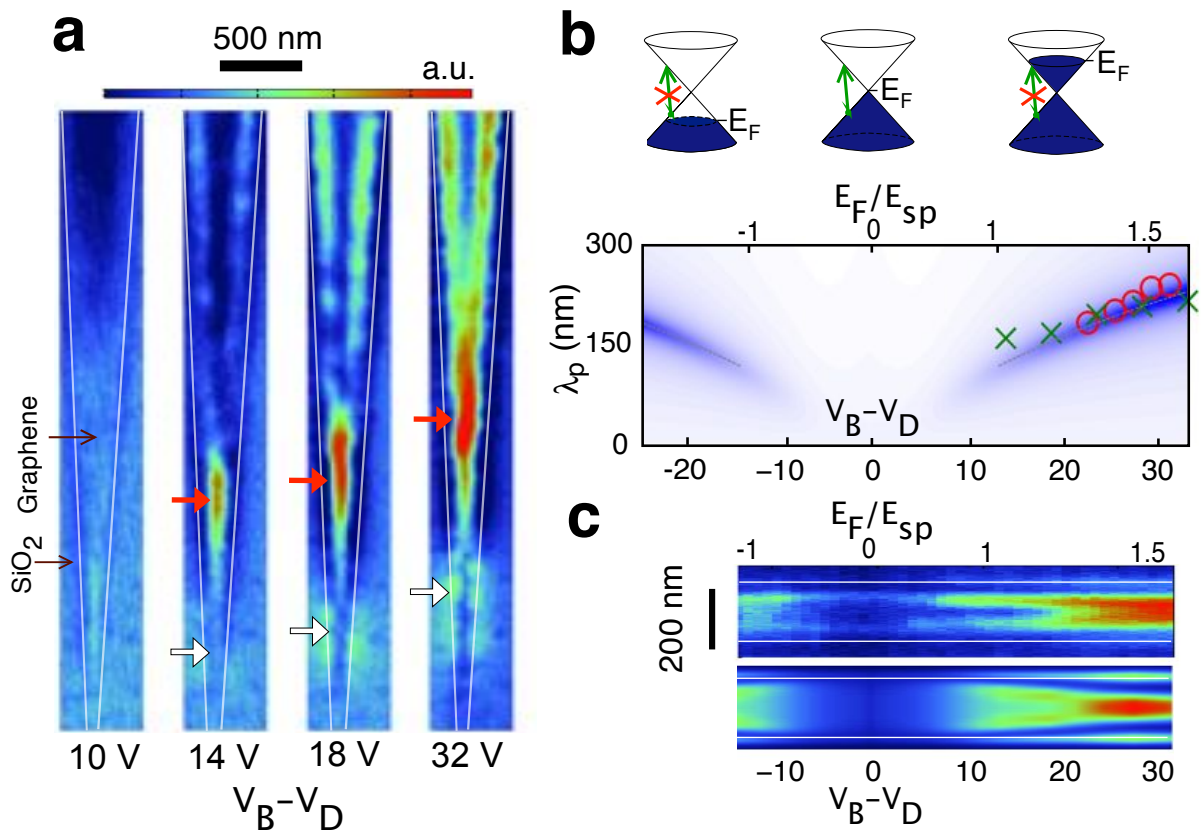

# Supplementary online material

belonging to the manuscript:

# Optical nano-imaging of gate-tuneable graphene plasmons

## Contents

## 1. Graphene plasmon dispersion

The graphene plasmon dispersion presented in the main text is based on the graphene conductivity within the random-phase approximation (RPA) in the local limit (i.e., for zero wave vector) which is given by [1]:

$$\sigma(\omega) = \frac{e^2 E_F}{\pi \hbar^2} \frac{i}{\omega + i\tau^{-1}} + \frac{e^2}{4\hbar} \left[ \theta(\hbar\omega - 2E_F) + \frac{i}{\pi} \log \left| \frac{\hbar\omega - 2E_F}{\hbar\omega + 2E_F} \right| \right],$$

where $E_F$ is the graphene Fermi energy and $\tau$ the relaxation time. The first term in this expression describes the conductivity associated with intra-band transitions, and it basically follows the semiclassical Drude model, while the second one describes inter-band transitions and becomes important only for high energies (i.e., for $\hbar\omega \gtrsim 2E_F$), giving rise to the observed 2.3% optical absorption. In our calculations, we use a conductivity similar to the above expression, modified to account for the effect of temperature [1], which we set to 300 K throughout this paper. We refer to it as the local RPA conductivity.

In the high doping regime, $E_F \gtrsim \hbar\omega$, where plasmons are long-lived excitations, the second term can be neglected, leading to equation (1) of the main text. In the region of interest, the plasmon dispersion relations obtained from this approximation and from the temperature-corrected local RPA are in good agreement, as we show in Figure S1.

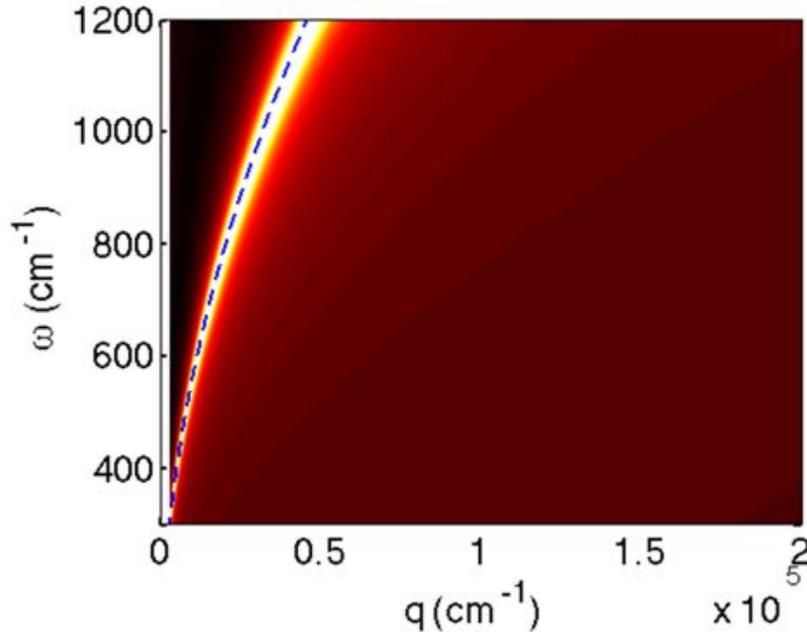

*Figure S1. Contour plot: plasmon dispersion calculated in the local RPA. Dashed line: plasmon dispersion given by equation (1) of the main paper. Mobility=10000, $E_F$=0.4eV, substrate permittivity $\varepsilon_r$=2.*

## 2. Graphene plasmons on a SiC substrate

The graphene plasmon dispersion is strongly affected by the dielectric function of the substrate $\varepsilon_r$. In Figure S2 we show the plasmon dispersion for graphene when the substrate is SiC. The SiC dielectric function includes phonon-polariton and plasmon-polariton terms and admits a double Lorentzian analytical approximation as [2]

$$\varepsilon(\omega) = \varepsilon_\infty \left(1 + \frac{\omega_{LO}^2 - \omega_{TO}^2}{\omega_{TO}^2 - \omega^2 - i\omega\Gamma} + \frac{\omega_p^2}{-\omega^2 - i\omega\gamma}\right).$$

As we can see from Figure S2, the plasmon dispersion is split into two branches due to interaction with the optical phonons of the substrate.

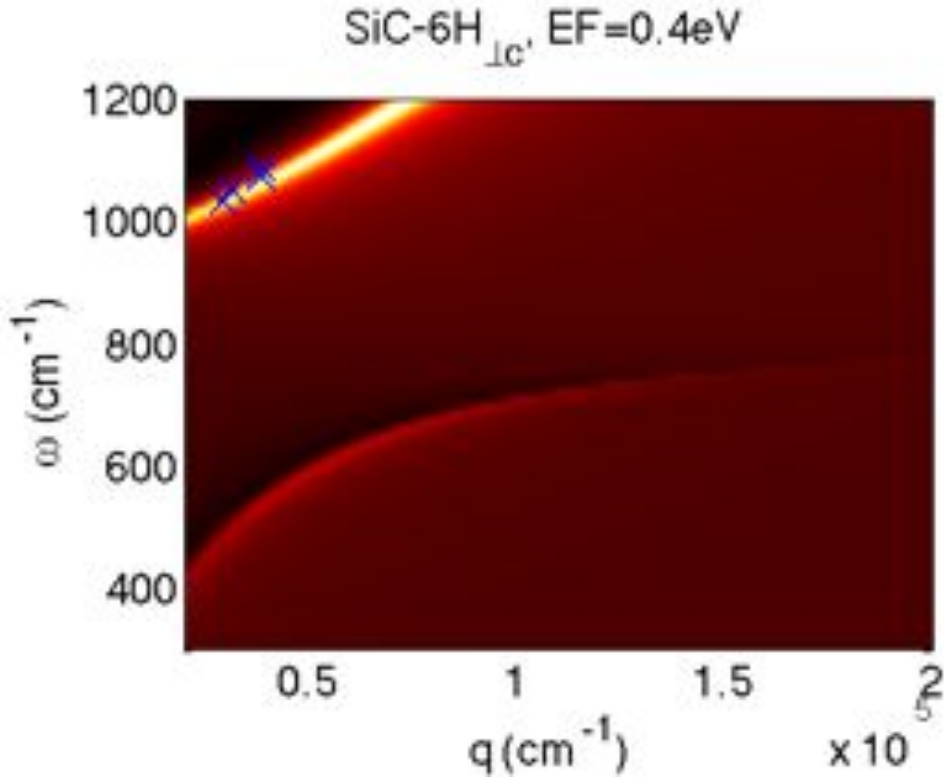

*Figure S2. Plasmon dispersion based on the local RPA for $\varepsilon_r$ corresponding to SiC (6H-SiC $\perp c$), with parameters [2] $\varepsilon_\infty$=6.56; $\omega_{TO}$=797 cm$^{-1}$ ; $\omega_{LO}$ =970 cm$^{-1}$ $\Gamma$ =5.9 cm$^{-1}$; $\omega_p$=230cm$^{-1}$; $\gamma$ =500cm$^{-1}$. Blue crosses: experimental data from Figure 2 of the main text.*

## 3. Sample preparation: SiC substrate

All samples were 1x1 cm$^2$ templates cut from a 3-inch, on-axis, semi-insulating C-face 6H-SiC wafer from Cree Research. Before cutting, electrochemical polishing was done by Novasic to get Epiready® morphology. A sacrificial oxide was then thermally grown, and chemically etched in HF to remove any (small) sub-surface damage from the polishing process. The necessary chemical treatments were clean-room compatible and very similar to the ones used before thermal oxidation or post-implantation annealing in standard SiC technology. Atomically flat surfaces were obtained in this way (no hydrogen etching was performed in our case to prepare the SiC surface).

The graphene growth was carried out in a RF furnace, in a secondary vacuum ($10^{-6}$ Torr). The first step of the growth was to heat the sample at 1150°C for 10 min in order to remove any trace of native oxide. The second step was to heat the sample at 1700°C for 30 min. During the growth, we artificially increased the C and Si partial pressures near the SiC surface by covering the sample with a graphite cap. This yields very long isolated graphene islands up to 300 μm long and 5 μm wide, surrounded by SiC, sparsely covering the surface [3,4].

The nucleation sites, often found at the center of long ribbons, are dislocations at the SiC surface, scratches or small particles. The ribbons are laying on wide atomically flat SiC terraces. These terraces are formed during the annealing process. This step-bunching phenomenon is standard in SiC technology, originating in a small initial miscut of the wafer surface with respect to the nominal 6H-SiC surface.

The initial SiC step edges, in the range of 2 nm high, prevent the graphene from growing perpendicular to the terraces. As a result, the graphene layer expands preferentially only on one terrace. The graphene seeds expand only along the terraces, which explains the significant length and shape of the ribbons. Let us note that not all of the graphitic long ribbons are monolayers. Depending on the size and the nature of the nucleating site, the ribbons are sometimes stacks of two or three layers of single layer graphene. The intrinsic doping due to substrate interactions are induced predominantly in the first layer, while the other layers are weakly doped.

Figures S3 and S4 show optical and Raman analysis of the sample used in the measurements. Figure S5 shows an optical image of a ribbon on SiC. The absorption measured from the optical contrast (on the right) is 1.5±0.3%, consistent with the expected 1.5% for single layer graphene on SiC. In Figure S4 a typical Raman spectrum of a graphene ribbon on SiC is displayed.

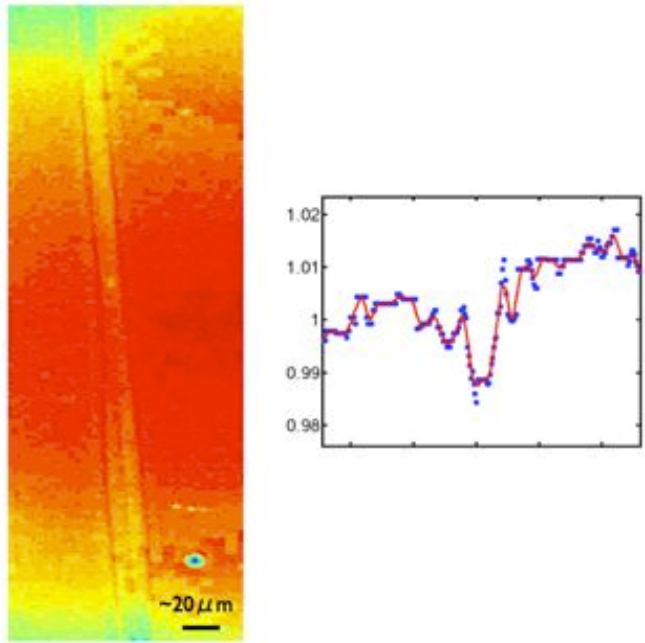

*Figure S3. Left: false-color optical image of a typical graphene ribbon on SiC. Right: optical contrast.*

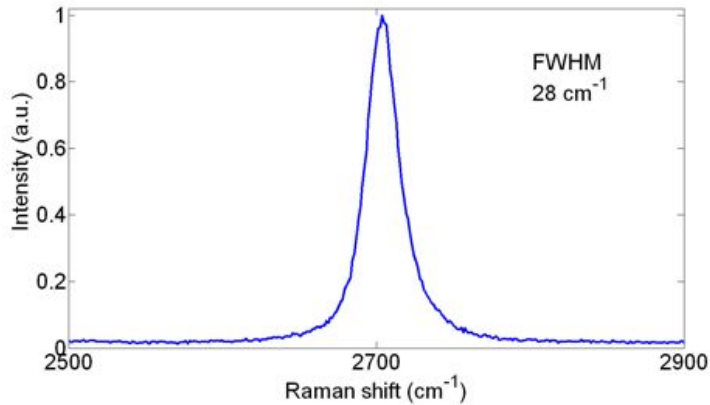

*Figure S4. Typical Raman spectrum (2D peak) of a graphene ribbon on SiC.*

**4. Sample preparation: SiO$_2$/Si substrate**

In order to tune graphene plasmons by an electrostatic gate, both CVD and mechanically exfoliated graphene were deposited on a SiO$_2$/Si substrate (285 nm SiO$_2$). As it concerns the CVD graphene films, monolayer graphene was synthesized using methane as the precursor gas and copper foil (Alfa Aesar) as the metal catalyst in a cold walled CVD reactor. The thickness of copper foil used was 25 μm and the foils were annealed using a hydrogen/argon atmosphere at 1000°C prior to the growth stage. The growth was performed at 1000°C using a low methane flow and 0.8 mbar. Once the growth was complete the graphene was transferred onto the required substrates via a wet transfer process [5]. In order to etch the copper, the graphene was first protected with a sacrificial polymethyl methacrylate (PMMA) layer. The etching solution used was a ferric chloride solution. Once the etching was complete the graphene was washed and transferred onto the SiO2 (285nm)/Si substrate. Finally, the PMMA layer was removed via thermal treatment.

Raman spectroscopy was used to characterize both types of samples, using a Renishaw Invia Raman Microscope (Fig. S5).

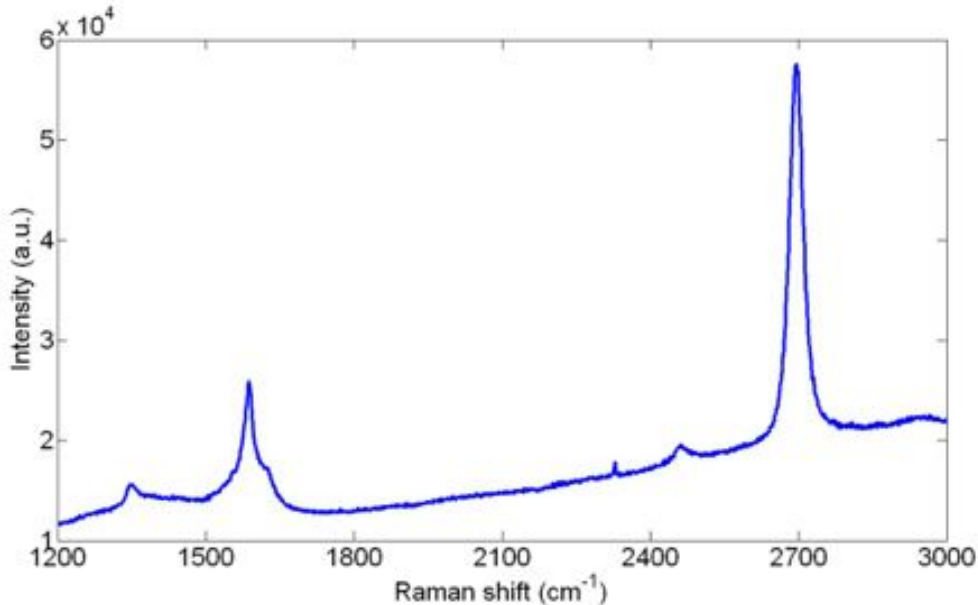

*Figure S5. Raman spectrum of a typical tapered ribbon based on CVD graphene, deposited on SiO$_2$ substrates. The 2D-peak has a FWHM of 37 cm$^{-1}$.*

Tapered ribbons were patterned on graphene by means of O$_2$ reactive ion etching of the areas previously defined with electron beam lithography (EBL) of polymethylmethacrylate (PMMA) resist. Contacts were then added to graphene with a second step of EBL with PMMA and subsequent evaporation of titanium and gold (5nm/100nm).

## 5. s-SNOM measurements and tip-graphene interaction

The measured s-SNOM signal in this work is proportional to the light intensity backscattered by the tip. The images were obtained in tapping mode, in which the distance between the sample and the tip oscillates with a typical amplitude of 50 nm, with an average distance of ~60 nm. The amplitude of the backscattered field also oscillates at the tapping frequency. From this signal, we extract the 1$^{st}$, 2$^{nd}$, 3$^{rd}$ and 4$^{th}$ harmonics. Higher harmonics in the optical signal are generated because the near-field signal depends nonlinearly on the distance between tip and sample. The data presented in the main text are obtained from the 3$^{rd}$ harmonic. This demodulation technique suppresses background signals, thus improving the signal-to-noise ratio.

The tip can potentially modify the graphene plasmon modes, so that one should actually consider the modes of the combined tip-graphene system instead of the modes launched by a point dipole (used for the LDOS calculations). This modification is expected to be weak if the tip-graphene distance is larger than the penetration distance of the plasmons in the $z$-direction. We find a penetration distance of ~16-32 nm (see below), which has to be compared to the tip-graphene distance (~60 nm on average during the measurements). The field intensity of the excited plasmons is therefore decaying by a sizeable factor at the position of the tip, and therefore, the feedback of the tip on the graphene plasmons is expected to be negligible.

The in-plane wave vector of graphene plasmons $2\pi/\lambda_s$ is ~100 times larger than the free-space wave vector $2\pi/\lambda_0$. This implies that the wave vector of the electromagnetic field associated with the plasmons must have components perpendicular to the graphene of the order of $k_\perp = 2\pi[1/\lambda_0^2 - 1/\lambda_p^2]^{1/2} \approx 2\pi i/\lambda_s$, which is nearly imaginary. This has important consequences, as the penetration distance of the fields into the surrounding medium along the direction $z$ away from the graphene is determined by the spatial dependence $\exp(ik_\perp z)$. The penetration distance (1/$e$ decay in intensity)) is thus $\sim 1/2\text{Im}\{k_\perp\} = \lambda_p/4\pi$. For the plasmon wavelengths measured in this paper, $\lambda_p$~200-400 nm, the penetration distance is ~16-32 nm. The fields emerging from graphene edges exhibit an even steeper decay away from the graphene, in order to compensate for the faster variation of the near field around the edges, which requires even larger in-plane wave vectors.

### 6. Extraction of plasmon wavelengths from the s-SNOM measurements

We retrieve the plasmon wavelength from the experimentally observed interference fringes. We use the fringes of the experimental data to extract $\lambda_p$, by drawing lines through maxima of the electric field. For each line, we start with an initial guess of the starting and final points; we then change the position and angle of the line, while monitoring the sum of the electric field intensity along the line. We optimize for the maximum value of the sum and plot the resulting line (Fig. S6). The wavelength is obtained by measuring the spacing between the lines, which is equal to half the wavelength, or if the lines are not perfectly parallel then the spacing between the end points is taken. Due to the curvature of the taper, lines are not perfectly parallel in some cases. However, this effect is taken into account in the error margin on the obtained wavelength. This procedure is shown in Figure S6 for a representative ribbon and different photon frequencies. We find significantly larger spacing between the two fringes at the edge, compared to the fringe spacing inside the ribbon. Qualitatively, this is reproduced in our LDOS calculations although the magnitude of the effect is stronger in the experiment.

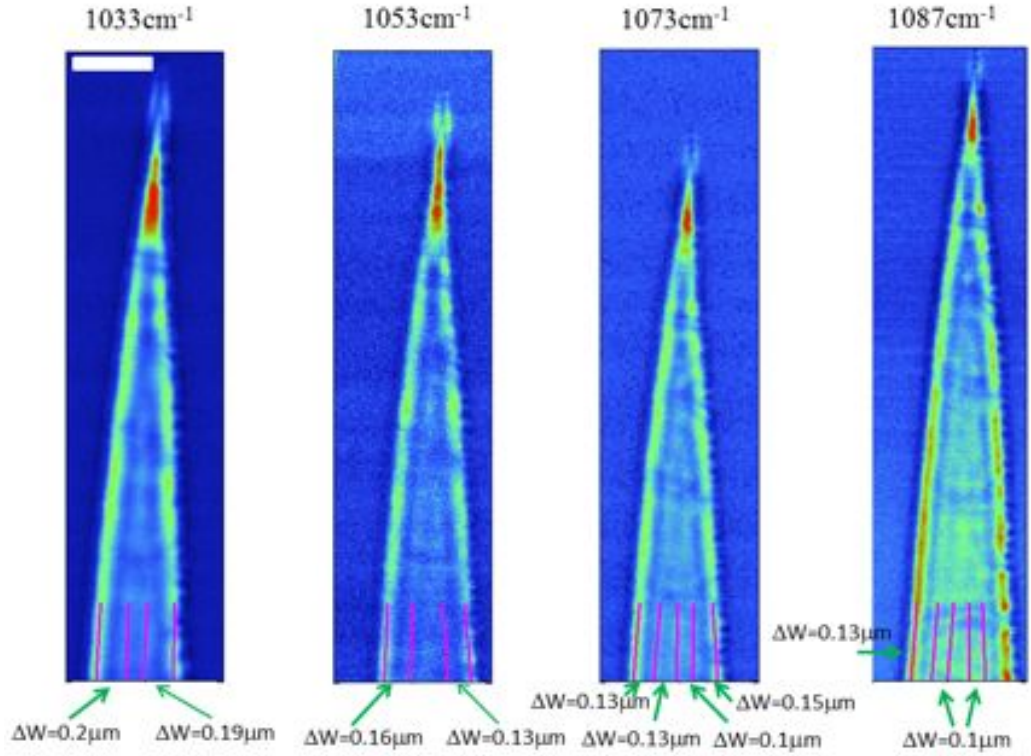

*Figure S6. 3$^{rd}$ harmonic near-field signal (arb. units) from graphene on SiC for different incident light frequencies. The lower pink lines show a fitting of the near-field fringe maxima used to derive the plasmon wavelength. ΔW distances are in µm.*

## 7. Calculations of the local density of optical states (LDOS)

The local density of optical states (LDOS) is defined, by analogy to the local density of states used to characterize electrons in solids, as

$$\text{LDOS} = \sum_j |\mathbf{E}_j(\mathbf{r}) \cdot \hat{\mathbf{n}}|^2 \, \delta(\omega - \omega_j), \tag{1}$$

where the sum runs over photon states *j* of energies $\omega_j$ and normalized electric fields $\mathbf{E}_j(\mathbf{r})$ [7], and $\hat{\mathbf{n}}$ is a unit vector along a selected direction. In contrast to their electronic counterpart, photons are vectorial quantities and their electric field needs a direction on which to project the LDOS. The LDOS clearly depends on the spatial position **r**, the frequency $\omega$ and the selected direction $\hat{\mathbf{n}}$. The LDOS can also be expressed as

$$\text{LDOS} = \frac{-2\omega}{\pi} \text{Im}\{\hat{\mathbf{n}} \cdot G(\mathbf{r},\mathbf{r}) \cdot \hat{\mathbf{n}}\}, \tag{2}$$

where *G* is the Maxwell Green tensor defined by

$$\nabla \times \nabla \times G(\mathbf{r},\mathbf{r}') - (\omega^2/c^2)\epsilon(\mathbf{r},\omega)G(\mathbf{r},\mathbf{r}') = \frac{-1}{c^2}\delta(\mathbf{r}-\mathbf{r}'). \tag{3}$$

This relation is easily verified in vacuum, where the electric fields satisfy

$$\nabla \times \nabla \times \mathbf{E}_j(\mathbf{r}) - (\omega_j^2/c^2)\mathbf{E}_j(\mathbf{r}) = 0$$

and the Green function reduces to

$$G(\mathbf{r},\mathbf{r}') = \sum_j \frac{\mathbf{E}_j(\mathbf{r}) \otimes \mathbf{E}_j^*(\mathbf{r}')}{\omega^2 - \omega_j^2 + i0^+}, \tag{4}$$

which is a solution of Eq. (3) due to the closure relation $\sum_j \mathbf{E}_j(\mathbf{r}) \otimes \mathbf{E}_j^*(\mathbf{r}') = \delta(\mathbf{r}-\mathbf{r}')$. Then, Eq. (1) follows upon insertion of Eq. (4) into Eq. (2), using the relation Im$\{(x+i0^+)^{-1}\}=-\pi\delta(x)$. Also in vacuum, we have

$$G(\mathbf{r},\mathbf{r}') = \frac{-e^{i\omega R/c}}{4\pi\omega^2 R^3}\left\{(\omega R/c)^2 + i\omega R/c - 1 - [(\omega R/c)^2 + 3i\omega R/c - 3]\frac{\mathbf{R}\otimes\mathbf{R}}{R^2}\right\}$$

where $\mathbf{R}=\mathbf{r}-\mathbf{r}'$, and the projected LDOS reduces to LDOS$_{vac}=\omega^2/3\pi^2 c^3$.

By construction, the Green function can be used to obtain the electric field produced by an external current distribution $\mathbf{j}(\mathbf{r})$ as

$$\mathbf{E}(\mathbf{r}) = -4\pi i\omega \int d\mathbf{r}' G(\mathbf{r},\mathbf{r}')\cdot\mathbf{j}(\mathbf{r}').$$

In particular, a point dipole $\mathbf{p}$ at $\mathbf{r}_0$ is equivalent to a current distribution $\mathbf{j}(\mathbf{r})=-i\omega\mathbf{p}\delta(\mathbf{r}-\mathbf{r}_0)$, and this allows us to obtain the LDOS in practice by solving the Maxwell equations for a dipole source. The LDOS at $\mathbf{r}_0$ then reduces to

$$\text{LDOS} = \text{LDOS}_{vac} + \frac{1}{2\pi^2\omega|p|^2}\text{Im}\{\mathbf{E}^{\text{ref}}(\mathbf{r}_0)\cdot\mathbf{p}^*\},$$

where $\mathbf{E}^{\text{ref}}$ is the field reflected by nearby structures and evaluated at the position of the source dipole. This is the procedure we actually follow to obtain the LDOS in this work, and $\mathbf{E}^{\text{ref}}$ is calculated by means of the boundary-element method (BEM) for a dipole source [8]. Additionally, the LDOS is connected to the decay rate of an excited emitter $\Gamma$ through $\Gamma=(4\pi^2\omega|d|^2/\hbar) \times$ LDOS, where $d$ is the excitation dipole strength.

In our experiment, the tip has an elongated shape along the direction *z* perpendicular to the graphene, with an apex rounding radius ~20 nm, which is sharp when compared to the light wavelength ~10 μm. Thus, the tip can produce an induced electric dipole mainly oriented along the *z* direction when illuminated by external light with a significant projection of the incident field on *z*. This allows us to assimilate the tip under external illumination to a point dipole oriented perpendicularly with respect to the surface. This dipole induces a plasmon that is scattered by the graphene edges, thus producing a *reflected* field $\mathbf{E}^{\text{ref}}$. This field is in turn scattered by the tip, and again, the scattered intensity is mainly sensitive to the *z* component, as the tip is comparatively much less efficient in producing scattering of fields oriented along the other two remaining directions. Therefore, the s-SNOM setup is collecting a complex amplitude (including information on the phase) that is proportional to the $E_z^{ref}$ component of the field produced by a dipole along *z* at its own position. The imaginary part of this field is the LDOS. It should be noted that when a mode is dominant in the LDOS, as is the case in the graphene ribbons for resonant widths, the LDOS is also proportional to $\left|E_z^{ref}\right|^2$ for that particular mode, as one can see from Eq. (1).

# 8. Analysis of the calculated LDOS, damping and comparison with particle in a box models

## 8.1 Tip distance

Figure S7 shows mode plots for different ribbon widths from LDOS calculations. We find best agreement between data and LDOS maps for a tip distance of 60 nm. As we can see, the LDOS show maxima at the edges of the ribbons, but they fall off super-exponentially while the LDOS inside the ribbon falls off exponentially. This may explain why the signal at the edges is small in the experimentally obtained near-field images.

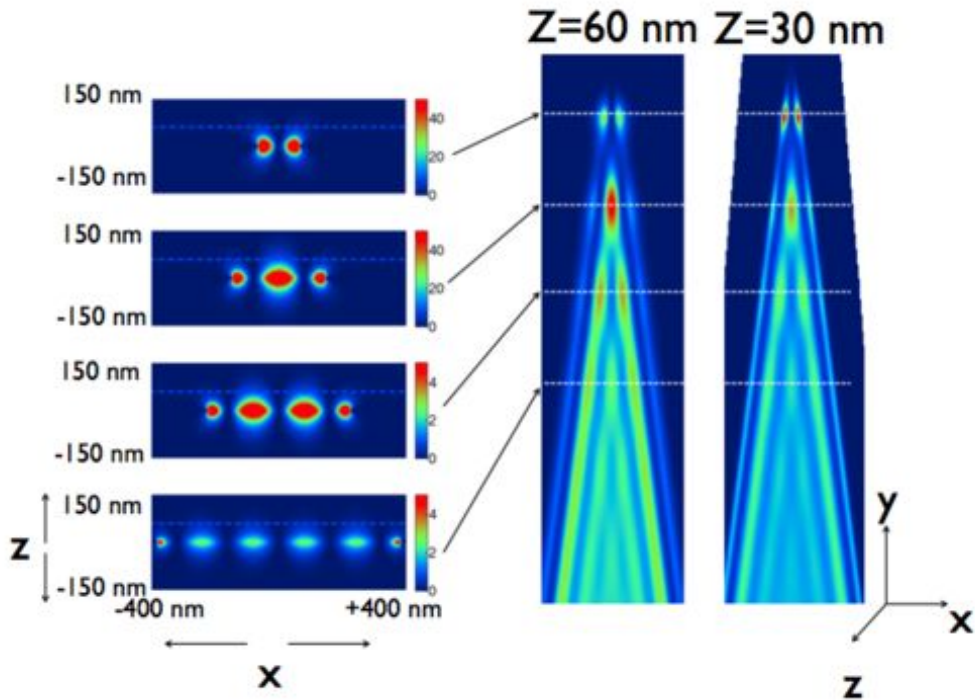

*Figure S7. Left: vertical sections of the calculated LDOS at different ribbon widths. Blue dashed line at 60nm, which is the average tip-graphene distance during the experiments. Right: LDOS for z=60 nm and z=30 nm, with $\lambda_0$=9.7 µm and $\varepsilon_r$=1.*

## 8.2 LDOS close to the graphene edges

Figure S8 shows the momentum distribution of the calculated LDOS as a function of position on a ribbon of width equal to 1000 nm. A clear peak in the wave vector distribution is observed inside the ribbon, explaining the equidistant LDOS fringes. At the edge, however, an additional wave vector peak can be observed, with a higher value than inside the ribbon. Therefore, the LDOS modulation at the edge displays a slightly longer period than inside the ribbon. This is qualitatively consistent with the experimental observations.

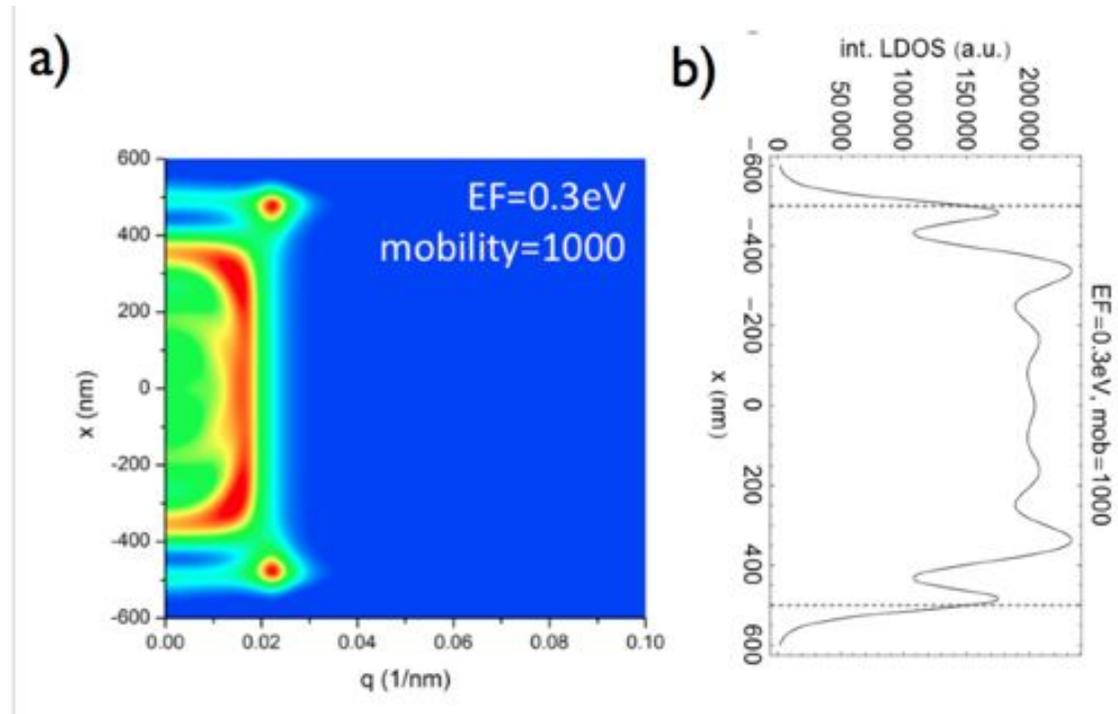

*Figure S8. a) LDOS wave vector distribution as a function of position on the ribbon. b) LDOS obtained by integration of the numerical data of a). Simulation parameters for a,b: dipole distance 60 nm, $\lambda_0$=9.7 μm and $\varepsilon_r$=1.*

*8.3 Plasmon damping*

The damping of the near-field signal away from the edges is partially due to the circular character of the plasmons excited by the tip. Additional damping is expected due to a combination of factors, such as substrate losses and intrinsic losses. We have incorporated intrinsic losses in the LDOS model by taking into account intra- and inter-band transition processes (see Sec. 1 above).

Fig. S9 shows LDOS linetraces for various tip distances. We observe a peak close to the edge (indicated by black arrow) with much larger intensity than the fringes inside the ribbon. The relative intensity depends strongly on the tip-distance. Therefore, we don't attribute this to plasmon losses but to electromagnetic field concentration at the edge.

In order to analyze intrinsitc plasmon losses, we examine the decay of the oscillations away from the higher-intensity peak. By comparing the decay of the experimentally observed oscillatons to the LDOS calculations, shown in Fig. S10, good agreement between measured data and theory is obtained for a mobility of 1200 $cm^2/Vs$. This analysis does not exclude other extrinsic or intrinsic loss mechanisms. More detailed experiments are required to elaborate on these aspects.

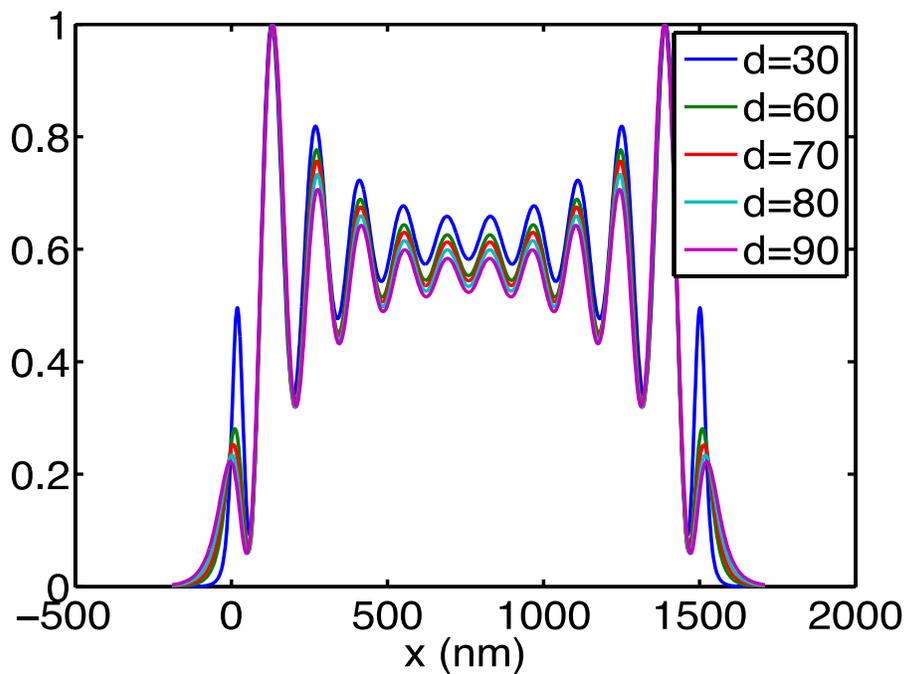

*Figure S9. Calculated LDOS for 5 different tip-distances.*

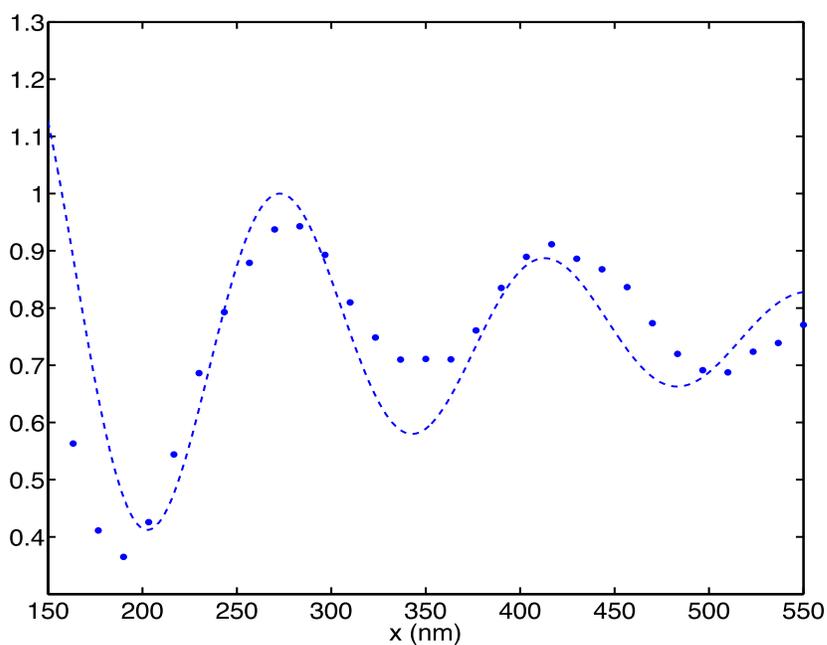

*Figure S10. Damping of the near-field signal away from the edge (dots, same as top left panel of Fig. 2 of main test), and calculated LDOS for a graphene ribbon with mobility $1200 cm^2/Vs$ (dashed line).*

### 8.4 LDOS profiles

In order to obtain a better physical understanding of the LDOS in a graphene ribbon, we compare in Fig. S11 the local density of optical states (LDOS) calculated for surface waves in a tapered ribbon for different boundary conditions.

In brief, we find that the calculated LDOS exhibits maxima close to the edges, similar to the experimental results. This is most clearly seen for the lowest order mode, near the tip of the ribbon. As we show below, our LDOS profiles are consistent with scalar surface waves with Neumann boundary conditions (zero-phase reflection), and in contrast to scalar waves with Dirichlet boundary conditions ($\pi$ phase reflection). Thus, graphene plasmons are being reflected at the boundaries with a 1 reflection coefficient, rather than the coefficient of −1 characteristic of the Fabry-Perot model. Our interpretation of graphene plasmons confined to a ribbon is equivalent to a Fabry-Perot model with zero-phase reflection coefficient. This situation is also encountered in plasmonic slot waveguides [10] and plasmonic nanoantennas [11], and it originates in the complexity of the reflection of 3D electric fields at the edges, which seems to be captured by effective Neumann boundary conditions in the scalar wave model.

Figure S11(a) shows the LDOS of graphene plasmons, as shown and discussed in the main text, obtained by solving Maxwell's equations for ribbons under the homogeneous ribbon approximation. This approximation consists of obtaining the LDOS for each point of the ribbon from a homogeneous ribbon with the same width as the local width of the tapered ribbon at the position of the point under consideration. The results for graphene plasmons are compared to the LDOS associated with scalar surface waves described by the Helmhotz equation and completely confined to the ribbon area. We present four different calculations for scalar waves: Figs. S11(b,c) correspond to Neumann boundary conditions (i.e., the scalar-wave amplitude is taken to have vanishing normal derivative at the ribbon edges, or equivalently, the reflection coefficient at the boundary is 1), whereas Figs. S11(d,e) are obtained using Dirichlet boundary conditions (i.e., the scalar wave amplitude vanishes at the boundary, which corresponds to a reflection coefficient equal to −1, similar to the quantum problem of a particle in a box with zero potential inside and infinite potential outside). The homogeneous ribbon approximation is used in Figs. S11(c,e), whereas Figs. S11(b,d) are calculated by solving the 2D wave for the actual tapered ribbon.

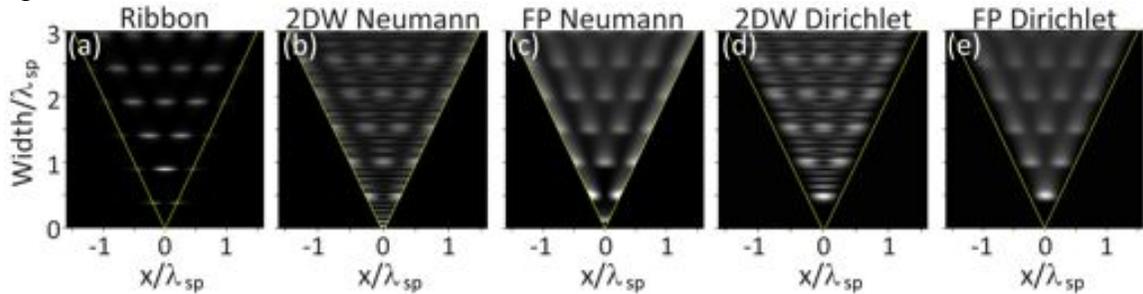

*Figure S11. Local density of states for different types of 2D surface waves confined to a tapered ribbon. (a) Graphene plasmons described by Maxwell's equations in the homogeneous ribbon approximation. The graph shows the local density of optical states (LDOS) projected along the graphene normal. (b-e) Scalar waves in an infinite potential well with Neumann's (b,c) and Dirichlet (d,e) boundary conditions, as calculated for an*

*actual tapered ribbon (b,d) and in the homogeneous ribbon approximation (c,e).*

The LDOS for graphene is obtained by solving Maxwell's equations as explained above. The LDOS is then projected along the graphene normal direction, which is the one probed in the experiment by the s-SNOM tip. For scalar waves in the actual tapered ribbon, we use a scalar version of the boundary-element method, whereby a point source is reflected at the boundaries and the reflected waves are used to obtain the local density of states. The reflected waves are represented in terms of self-consistent boundary charges. Further details about this method can be found elsewhere [9]. In practice, we consider finite tapered ribbons extending four times the propagation distance beyond the region shown in Fig. S11, so that spurious reflections at the upper boundary cannot reach the region of the images.

We remark that the LDOS is dominated by resonances at specific positions along the ribbon. This is best illustrated by the calculation for surface waves under Dirichlet boundary conditions in the homogeneous ribbon approximation, which can be carried out analytically. For a surface wave vector $k=2\pi/\lambda$, determined by the wavelength $\lambda$, the surface states corresponding to a given ribbon width $W$ are $N_{n,k_\parallel} \cos(n\pi x/W) \exp(ik_\parallel y)$, with $n=1, 2, ...$, where $N_{n,k_\parallel}$ is a normalization constant, $x$ extends from $-W/2$ to $W/2$ across the ribbon, and $y$ is the coordinate along the long ribbon direction. Here, $k_\parallel = [k^2 - (n\pi/W)^2]^{1/2}$ is the parallel wave vector along $y$, and there is a finite number of modes for each value of $W$, determined by the condition that $k_\parallel$ is real, or equivalently, $n < kW/\pi$ (obviously, no mode is allowed for wavelengths $\lambda > 2W$). Using wave-vector $\delta$-function normalization for the states, the normalization coefficient satisfies $N_{n,k_\parallel} = (k/\pi k_\parallel W)^{1/2}$. Therefore, each state contributes to the local density of states with a term $N_{n,k_\parallel}^2 |\cos(n\pi x/W)|^2$, which has a $1/k_\parallel$ divergence when a new mode becomes allowed as $W$ is increased. This divergence is actually broadened by the finite surface-wave propagation distance $L$, introduced through an imaginary part of the wave vector $\text{Im}\{k\}=1/2L$. As a result, we obtain broad maxima (see Fig. S11(e)), with $n=1, 2, ...$ anti-nodes per state, consecutively showing up as the ribbon width increases. A similar explanation applies to all kinds of surface waves considered in Fig. S9. In particular, for Neumann boundary conditions, the modes have $\sin(n\pi x/W)$ profiles.

The validity of the homogeneous ribbon approximation is clearly established by comparing Figs. S11(b) and S11(c) for Neumann boundary conditions. These plots show a similar structure of broad maxima at the positions expected from the above analysis. Finer structure in the local density of states is caused by reflection at the ribbon apex, emerging as smaller, weaker features. Likewise, Fig. S11(d) compares reasonably well with Fig. S11(e), thus validating the homogeneous ribbon approximation for Dirichlet boundary conditions as well. These results are reassuring, as we are presenting electromagnetic calculations under this approximation for graphene, because tapered ribbons with the dimensions considered here are currently computationally too demanding to be performed in a reasonable time.

## 9. Additional experimental data

In the following figures we show additional data of the s-SNOM measurements performed on tapered graphene ribbons on SiO$_2$/Si. In Fig S12, the data corresponding to the green crosses in Fig. 4b of the main paper are displayed, while Fig. S13 contains the data used for Fig. 4c. Fig. S14 contains near-field data from a bilayer exfoliated graphene flake, revealing the plasmon response for both p and n-type charge carriers.

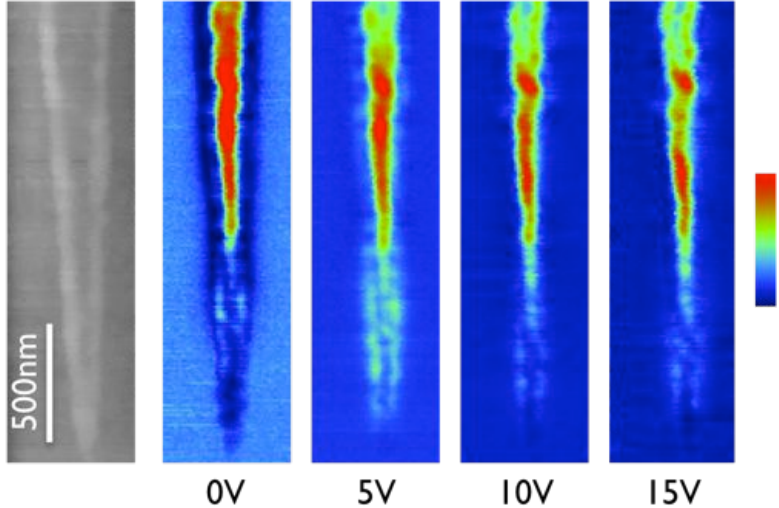

*Figure S12. Topography (left) and near-field amplitude images for different backgate voltages from the sample represented with green crosses in Fig. 4b of the main paper.*

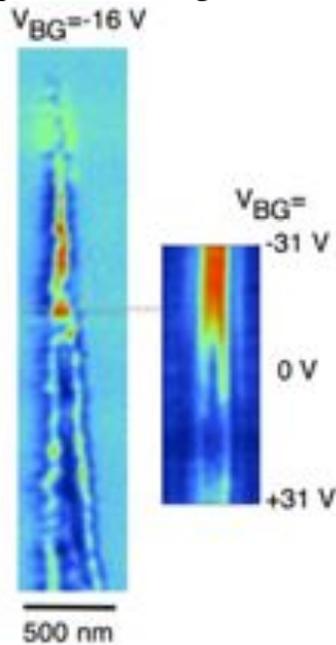

*Figure S13. Left: full near-field image for a ribbon used to plot Fig 4c of the main paper. Right: near-field amplitude signal obtained by measuring along the dashed line while sweeping the backgate voltage.*

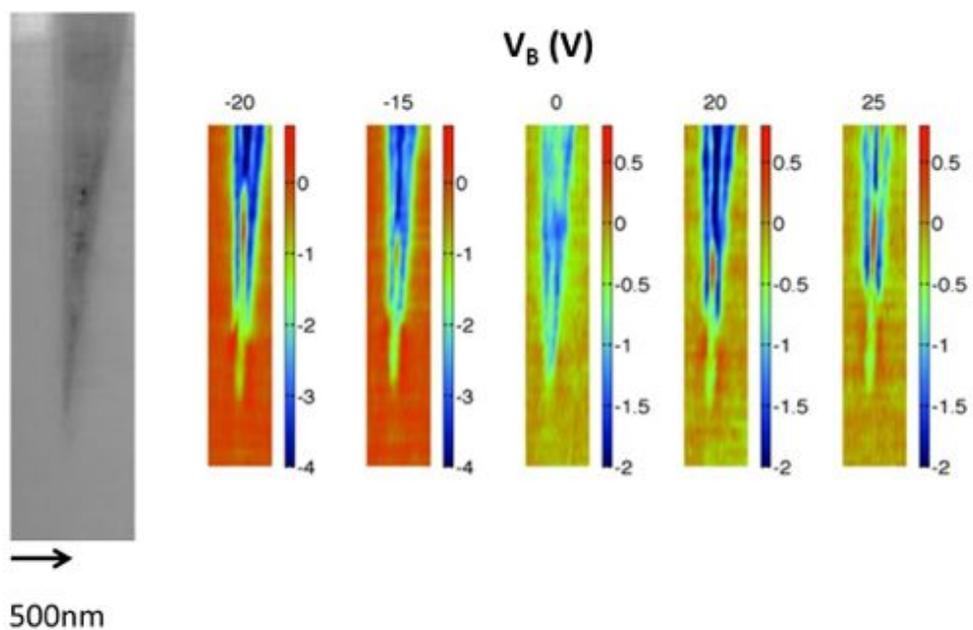

*Figure S14. Topography (left) and near-field amplitude images for different backgate voltages from the exfoliated bilayer graphene flake (not shown in main text).*

# References

1. A Falkovsky. Optical properties of graphene, *J. Phys.: Conf. Ser.* 129 012004 (2008).
2. Hofmann, M., Zywietz, A. & Karch, K. Lattice dynamics of SiC polytypes within the bond-charge model. Phys. Rev. B (1994).
3. N. Camara, J.-R. Huntzinger, G. Rius, A. Tiberj, N. Mestres, F. Pérez-Murano, P. Godignon, and J. Camassel. Anisotropic growth of long isolated graphene ribbons on the C face of graphite-capped 6H-SiC, *Phys. Rev. B* **80**, 125410 (2009).
4. N. Camara, G. Rius, J.-R. Huntzinger, A. Tiberj, L. Magaud, N. Mestres, P. Godignon, and J. Camassel. Early stage formation of graphene on the C face of 6H-SiC, *Appl. Phys. Lett*. **93**, 263102 (2008).
5. X. Li, W. Cai, J. An, S. Kim, J. Nah, D. Yang, R. Piner, A. Velamakanni, I. Jung, E. Tutuc, S.K. Banerjee, L. Colombo and R.S. Ruoff *Science* **324**, 1312, 2009
6. Novoselov, K. S. *et al.* Two dimensional atomic crystals. *Proc. Natl Acad. Sci. USA* **102**, 10451–10453 (2005).
7. R. J. Glauber and M. Lewenstein, Phys. Rev. A 43, 467 (1991).
8. F. J. García de Abajo and A. Howie, Phys. Rev. B **65**, 115418 (2002).
9. F. J. García de Abajo, J. Cordón, M. Corso, F. Schiller, and J. E. Ortega. Lateral engineering of surface states – towards surface-state nanoelectronics, *Nanoscale,* **2**, 717-721 (2010).
10. Barnard, E. S., Coenen, T., Vesseur, E. J. R., Polman, A. & Brongersma, M. L. Imaging the Hidden Modes of Ultrathin Plasmonic Strip Antennas by Cathodoluminescence. Nano Lett 11, 4265–4269 (2011).
11. Dorfmüller, J. et al. Fabry-Pérot Resonances in One-Dimensional Plasmonic Nanostructures. Nano Lett 9, 2372–2377 (2009).